\documentclass[runningheads]{llncs}
\usepackage[T1]{fontenc}

\usepackage{amsmath} 
\usepackage{amssymb}  

\usepackage{graphicx}
\usepackage{multirow}
\usepackage{booktabs} 

\usepackage{subfig}

\usepackage{microtype}

\usepackage{hyperref}

\def\subparagraph{}

\usepackage{titlesec}

\titlespacing*{\section}{0pt}{0.3cm}{0.2cm}
\titlespacing*{\subsection}{0pt}{0.3cm}{0.2cm}
\titlespacing*{\subsubsection}{0pt}{0.3cm}{0.2cm}

\begin{document}

\title{Wireless Interference Recognition with Limited Data via Utilizing Diffusion Model}
\titlerunning{Wireless Interference Recognition}
%
\author{Rui Wang\inst{1,2} \and Tong Li\inst{1,2} 
Chanchan Zhu*\inst{3} \and 
Shaolong Gu \inst{3} 
\and Zhu Xiao\inst{1,2}
}

\authorrunning{R. Wang et al.}
%
\institute{College of Computer Science and Electronic Engineering, Hunan University. \and
Ministry of Education Key Laboratory of ``Fusion Computing of Supercomputing and Artificial Intelligence'', Changsha, China. \and The 54th Research Institute of China Electronics Technology Group Corporation (CETC),
Shijiazhuang, Hebei 050081, China.\footnotetext{* Corresponding author}\\
\email{wray@hnu.edu.cn, tliay@hnu.edu.cn, chanchannibang1988@126.com, Zhxiao@hnu.edu.cn}\\
}

\maketitle              

\begin{abstract}
Accurate recognition of interference signals is essential for maintaining the reliability and security of modern wireless communication systems.
However, existing approaches face several significant challenges: the high cost and the scarcity of accurately labeled datasets adaptable to diverse environments, coupled with the limited ability to evaluate uncertainty factors arising from propagation in complex electromagnetic environments. To address these challenges, this paper proposes a diffusion-driven model for wireless interference recognition. Specifically, we develop a composite data augmentation strategy that integrates geometric transformation and modal augmentation to enrich the limited data distribution. Geometric transformation expands scarce signal data, and modal augmentation generates semantically consistent multimodal data by modeling the correlations among cross-domain features. Following that, we propose a novel diffusion-based generative model with an interference classifier to accurately capture the distribution of interference signals within given signal data. 
It hence significantly improves the understanding of environmental uncertainty factors. Extensive experiments demonstrate that the proposed model outperforms state-of-the-art methods by over 5.84\% in wireless interference recognition.

\end{abstract}

\section{Introduction}
\label{sec:intro}

As the torchbearer of next-generation telecommunications, sixth-generation (6G) networks are expected to evolve wireless communications networks into a new era defined by higher frequencies, massively increased capacity, and ultra-low latency~\cite{li2023carbon,add-3}. However, with the deep integration of smart devices into daily life and the increasingly complex electromagnetic propagation environment, wireless communication interference has become increasingly prominent, posing a significant threat to the security and reliability of these systems~\cite{xiao2025compound,add-1}. As such, enhancing the ability to recognize wireless interference is critical to ensuring the safe and stable operation of wireless communication systems and represents a fundamental requirement for improving the quality and efficiency of 6G networks~\cite{add-4,add-2}.

Existing work has focused on improving the accuracy of interference recognition. In earlier work, researchers used likelihood-based methods to construct test statistics and develop optimal judgment thresholds based on the Bayesian minimum error probability criterion to achieve accurate recognition results~\cite{intro-meth1}. 
On the downside, likelihood-based methods rely heavily on domain expert knowledge, limiting their applicability and scalability across a wide range of applications. 
Lately, deep learning (DL) has introduced a new research paradigm for accurate interference recognition due to its powerful feature extraction and data mining capabilities. For example, Yang et al.~\cite{intro-meth2} employed a lightweight residual network with dilated kernels to effectively enhance the performance of interference recognition. 
An et al.~\cite{intro-meth3} integrated soft shrinkage into the DL model, achieving superior interference recognition performance in complex channel-propagation environments. 
Although DL-based methods have made significant progress in predicting the mean label corresponding to input signals, they face two major issues. \textit{i}) DL methods remain limited in estimating uncertainty factors in diverse electromagnetic propagation environments. 
\textit{ii}) Most supervised learning approaches are heavily dependent on sufficient labeled data. When signal data is insufficient or imbalanced, DL models fail to achieve satisfactory recognition performance due to fuzzy classification boundaries.

Recently, generative AI has achieved remarkable performance thanks to its high-quality generative capabilities, flexibility, and scalability~\cite{li2025generative,su2025gennet,chai2024generative}. To address the uncertainty factors in various electromagnetic propagation environments, generative AI methods, such as diffusion models, have demonstrated  potential in interference recognition tasks involving numerous nonlinear and non-stationary factors, due to their ability to model underlying probability distributions and data dependencies~\cite{zhang2024netdiff}. To address insufficient and imbalanced data, geometric and modal transformations of limited signal data effectively expand the dataset, address data scarcity, and improve robustness and generalization, enabling outstanding performance in complex data-driven interference recognition tasks.

Motivated by this, we propose a \underline{Diff}usion-driven model for \underline{W}ireless 
\underline{I}nterference \underline{R}ecognition (\underline{DiffWIR}) under limited data conditions. Specifically, the proposed approach consists of the composite data augmentation module and the diffusion-based recognition module. In the composite data augmentation module, we expand a limited labeled dataset using various transformation techniques, including stretching, flipping, and adding noise. 
Besides, we employ modal augmentation techniques to conduct an in-depth analysis of the expanded dataset, aiming to extract richer, multidimensional signal representations.
By using the short-time Fourier transform (STFT) and the Fourier transform (FFT), we extract time-frequency and power spectral density images from the signal data, enriching data representation and providing stronger support for data-driven supervised model training. In the diffusion recognition module, we design a diffusion-based model for the interference recognition task by replacing continuous response variables with one-hot-encoded label vectors. By replacing the average estimator with the pre-trained classifier, the classifier outputs soft probabilities for different interference category labels. 
Furthermore, the classifier, combined with a Markov chain and Langevin sampling, inherently enhances the understanding of uncertainty factors in spatial propagation, thereby enabling accurate interference recognition under complex environmental conditions.

The contributions are summarized as follows.

$\bullet$ We propose a composite data augmentation method that allows supervised DL algorithms to operate efficiently  on limited datasets. By employing synergistic spatial and multi-domain augmentations, the proposed method not only alleviates overfitting under limited dataset conditions, but also enhances model robustness by capturing geometric invariance and cross-modal correlations, thereby facilitating more comprehensive feature learning.

$\bullet$ We propose a novel diffusion-based model to improve the robustness of wireless interference recognition. Specifically, we leverage the diffusion model to gradually recover data from noise and effectively capture the rich data distributions provided by the data augmentation module, thereby enhancing the perception of environmental information in spatial propagation. Such modeling capability enables the model to deliver more accurate interference recognition under complex propagation conditions.

$\bullet$ We conduct extensive experiments on datasets collected from real-world scenarios. The results demonstrate that the proposed model achieves approximately 5.84\% higher accuracy compared to the baseline model, underscoring its superior performance and effectiveness.

\section{Challenges}
The performance of traditional supervised deep learning algorithms is strongly related to the quality of the dataset. However, in limited datasets, insufficient and unbalanced samples can lead to model overfitting and poor generalization. Therefore, effectively expanding the amount of data and enriching the data distribution are key challenges for wireless interference recognition. In addition, the inherent uncertainty factors in wireless communication signal propagation (e.g., time-varying multipath fading, non-Gaussian noise coupling, and Doppler-induced distortion) may induce nonlinear overlaps in the high-dimensional feature space, resulting in degraded decision boundaries among interference classes. Uncertainty assessment in the context of such a complex interference recognition task represents another significant challenge that must be addressed.

\subsection{Chanllenge 1: Augmentation of Limited Datasets}
Most existing interference recognition architectures based on the supervised deep learning paradigm demonstrate excellent recognition performance when two essential prerequisites are met: \textit{i}) the availability of a class-balanced dataset with a large number of accurately labeled samples; \textit{ii}) a high degree of similarity between the data distributions of the trainset and the testset~\cite{chan-1}. Accurate data labeling requires domain experts, a process that is typically time-consuming and labor-intensive. Moreover, differences in signal transmitter configuration parameters and the complex electromagnetic environments in which data are collected result in significant variations in signal data distribution. Therefore, acquiring sufficient labeled datasets for diverse scenarios is often impractical. However, in limited datasets, data scarcity and imbalance often lead to issues such as model overfitting, reduced generalization, and insufficient learning of complex features. In response to such conditions, significant efforts have been devoted to improving the robustness of wireless interference recognition methods, including the design of more complex deep neural network structures, the development of advanced loss functions, and research into signal data augmentation techniques. Conventional geometric data augmentation strategies (e.g., flipping and random cropping) are restricted to the sample distribution domain of the original data feature space and can generate only a limited range of transformed samples. Their inherent limitations hinder their ability to effectively mitigate the problem of missing cross-modal features. In contrast, single-modality augmentation techniques (e.g., noise addition) lack the ability to model cross-domain feature correlations and struggle to produce semantically consistent multimodal data. Hence, collaboratively expanding data distributions across the spatial domain (via geometric transformations) and other signal-feature domains (via modal generation) using a composite data augmentation strategy is a key challenge in wireless interference recognition under limited data conditions.

\subsection{Challenge 2: Evaluation of Complex Channel Uncertainties}
Wireless signals propagating through dynamic channels are inherently plagued by non-stationary impairments such as time-varying multipath fading, Doppler shift, and non-Gaussian interference coupling~\cite{ref_jiang_unirm,jiang2024physics}. These impairments introduce random distortions to the signal representation, leading to overlapping feature distributions in high-dimensional space and degraded decision boundaries between interference classes. Traditional deep learning frameworks rely on deterministic predictions or global statistical priors, which cannot capture the instance-specific ambiguity caused by channel distortions. For example, frequency-selective fading and impulse noise in non-line-of-sight (NLOS) scenarios produce overlapping feature representations in the high-dimensional signal space, leading to overconfident interference misclassifications that cannot be evaluated probabilistically by traditional classifiers. The cumulative effect of these influencing factors propagates errors in the multi-stage interference recognition process, exacerbating the inaccuracy of the final recognition results. Integrating the uncertainty evaluation of complex propagation environments into the wireless interference recognition represents another critical challenge that must be addressed~\cite{chan-2}.

\begin{figure}[t]
    \centering
    \vspace{-3mm}
    \includegraphics[width=0.68\linewidth]{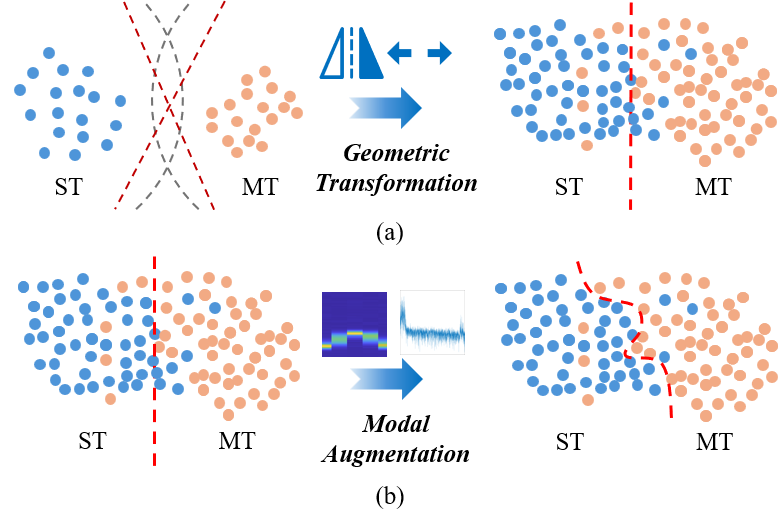}
    \vspace{-3mm}
    \caption{Composite data augmentation strategy for wireless interference recognition under limited data conditions. For example, consider the binary classification task of ST and MT (noting that the actual scenario is more complex). Geometric transformation compresses the feature space and pushes the classification boundary toward the optimal decision by expanding the volume of signal data. Modal augmentation refines the classification boundary by enriching the intrinsic information representation of the signal, allowing for better discrimination of highly similar samples that are prone to misclassification.}
    \vspace{-3mm}
    \label{fig:Aug}
\end{figure}

\begin{figure*}[t]
    \centering
    \vspace{-2mm}
    \includegraphics[width=0.9\linewidth]{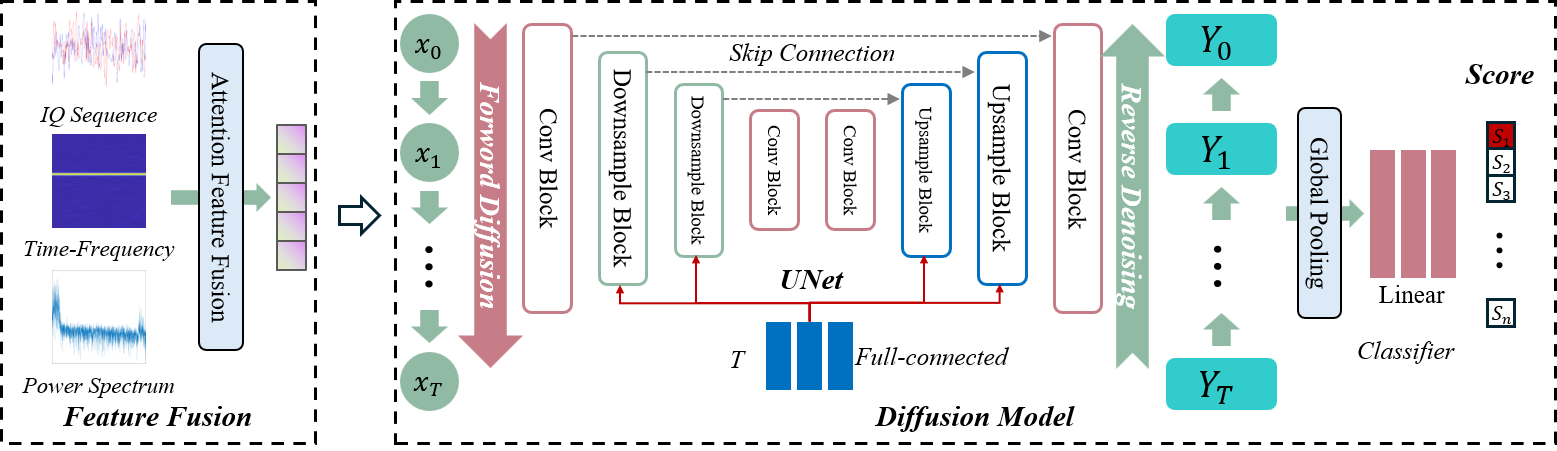}
     \vspace{-2mm}
    \caption{The detailed architecture of the diffusion-based recognition module. Multimodal features obtained through data augmentation are fed into the diffusion model following attention fusion. After the forward diffusion and reverse denoising processes, the resulting output is passed to the classifier to produce the interference recognition result.}
    \label{fig:Diff}
     \vspace{-2mm}
\end{figure*}

\section{Solution}\label{model}
The proposed method comprises two key modules: \textit{i)}~The composite data augmentation module that leverages geometric transformation and modal augmentation to enrich the sample distribution information for different types of disturbances; \textit{ii)}~The diffusion-based recognition module designed to generate confidence scores for disturbance recognition results and assess uncertainty using a diffusion model.

\subsection{Composite Data Augmentation Module}
Given the high cost of building datasets tailored to diverse electromagnetic environments, distinguishing different types of interference becomes challenging when data-driven deep learning algorithms fail to thoroughly analyze sample distribution information. To better enrich sample data and generate meaningful data representations, we develop the composite data augmentation module. Specifically, geometric transformations are applied to generate data sample variants, increasing the sample size, while multi-modal samples are generated via modal augmentation to supplement feature representations.

Inspired by popular data augmentation techniques in image processing, we develop a geometric transformation-based data augmentation method tailored for wireless IQ signals. Unlike transformations such as arbitrary rotation and flipping commonly used in image processing, IQ signals are sequential data with strong temporal correlations. Improper operations on signals can disrupt the coded modulation scheme, leading to meaningless instances. Therefore, we employ strategies such as flipping, stretching, and adding noise to individual components to expand the limited dataset. Fig.~\ref{fig:Aug}(a) illustrates the impact of geometric transformation-based dataset expansion on the interference recognition task. Initially, when data is limited, the algorithm has significant room for adjustment but struggles to converge gradually toward the optimal solution from suboptimal ones. As the amount of sample data increases, the distinctions between sample categories become clearer, and the decision boundary approaches the optimal.

Fuzzy samples with high similarity near the decision boundary cannot be divided into finer granularity. This is primarily because the model lacks the ability to represent complex features due to the limitation of single-modal data. To address this problem, we leverage STFT and FFT to generate time-frequency images and power spectral density (PSD) images for modal augmentation, enriching the internal representation of the sample data. The PSD describes the power distribution of a signal in the frequency domain using probabilistic statistical methods. Intuitively, it  represents the energy distribution of a signal at different frequencies by calculating the average energy at specific frequency points. However, the FFT cannot perform joint time-frequency analysis. To explore the correlation between the two dimensions of time and frequency, we employ a time-frequency analysis method based on STFT. By applying a sliding window function across the entire time axis, this method provides local information about the signal in both time and frequency domains and is particularly suitable for analyzing non-stationary signals in complex electromagnetic environments. As shown in Fig.~\ref{fig:Aug}(b), generating multimodal signal data further enhances the details of the classification boundary and significantly improves the discrimination of samples with high similarity near the boundary.

\subsection{Diffusion-based Recognition Module}
Unpredictable fluctuations caused by channel propagation in complex environments significantly degrade the accuracy of interference recognition. To address this issue, we propose a data-driven diffusion model recognition module, as illustrated in Fig.~\ref{fig:Diff}. Specifically, the diverse information representation provided by fusion modal augmentation serves as the output, while the diffusion model captures subtle spatial fluctuations by leveraging forward diffusion and backward denoising to achieve accurate wireless interference recognition.

Given the varying importance of information contained within multimodal data, it is essential to adaptively learn appropriate weights for different feature types. To meet this requirement, we employ the attention feature fusion mechanism to address the challenges posed by diverse structural features~\cite{fan2021understanding}. The main component of this mechanism is the multi-scale channel attention module, which extracts both global and local feature attention through two distinct branches. Lightweight point convolution is used to capture local feature attention, while the global channel attention mechanism incorporates global average pooling from the first layer of point convolution, better preserving the global and spatial information of the input signal features.

The core of the diffusion model is a novel generative framework inspired by the diffusion process in non-equilibrium thermodynamics. The model comprises a forward diffusion process and a backward denoising process. The forward diffusion process gradually adds Gaussian noise to the data, following a predetermined strategy (e.g., linear or cosine scheduling), transforming the input data into pure noise that approximately follows a normal distribution. The backward denoising process leverages a U-Net to learn the noise distribution, predict the noise or data distribution at each step, and iteratively denoise via the reverse process to gradually reconstruct the original data. UNet, as the core of noise prediction, features a symmetric encoder-decoder structure with skip connections. By encoding the diffusion steps as a vector (e.g., sinusoidal position encoding) and injecting it into each layer of the UNet, the model becomes aware of the current denoising stage. A self-attention or cross-attention layer inserted between the encoder and decoder enhances global modeling capabilities. To accomplish the final interference recognition task, we append a classification head to the diffusion model and design a joint loss function combining diffusion and classification to better capture environmental uncertainty.

\section{Evaluation}
\subsection{Datasets and Baselines}
\begin{figure}[tb]
    \centering
    \includegraphics[width=0.72\linewidth]{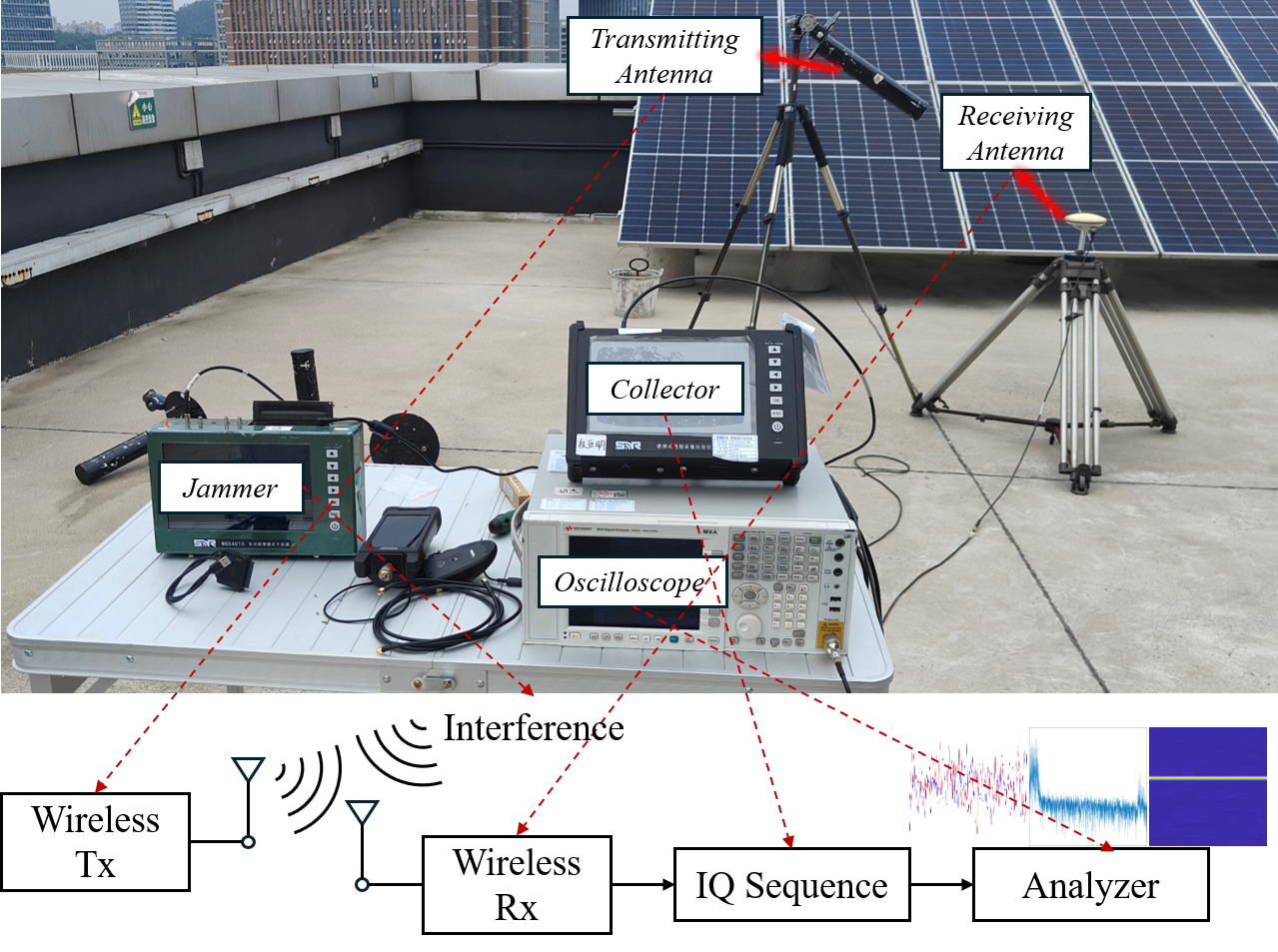}
    \caption{The experimental dataset is collected from real-world scenarios. 
    The jammer is utilized to simulate diverse interference. The data is ultimately collected and stored in binary format using hardware instruments. The oscilloscope can used to analyze interference in a real-time manner.}
    \label{fig:dataset}
\end{figure}

The dataset used in our experiment is derived from the wireless communication signal interference dataset collected in real-world experiments~\cite{add-2}. The deployment environment of the interference scenario is illustrated in Fig.~\ref{fig:dataset}. The portable wireless jammer is employed as a signal generator to transmit interference signals at the carrier frequency of 1268 MHz. The wireless jammer is designed to generate interference signals with various parameters. A portable collector is used to receive interference signals, with a sampling rate of 50 MHz. The transmission antenna is directional. Based on the above-described collection environment, we obtain 10 types of interference signals, including single-tone (ST), multi-tone (MT), pulse (PL), frequency modulation (FM), linear frequency modulation (LFM), noise frequency modulation (NFM), sinusoidal frequency modulation (SFM), tamed frequency modulation (TFM), noise amplitude modulation (NAM), and radio frequency noise (RFN). For different types of interference signals, we randomly sample to create a limited, unbalanced dataset of 100 signal samples. The test set contains 6,000 signal samples collected from interference signals with varying parameters, ensuring fairness and reliability in experimental verification. After data augmentation, we expand the trainset to the same order of magnitude as the testset and generate an equal number of time-frequency images and PSD images through conversion. To evaluate the method's interference recognition performance, we select overall accuracy (OA), recall rate (Recall), and the kappa coefficient (KA) as evaluation metrics. 

To evaluate the performance of the proposed  model, we compare it with several up-to-date methods.

\textbf{SENET \cite{baseline-se}.} SENET is a deep neural network architecture designed based on the channel attention mechanism. Its core innovation lies in dynamically modeling feature-channel dependencies using the squeeze-and-excitation module, thereby adaptively calibrating channel feature responses. This network can significantly enhance the model's representation ability in interference recognition tasks and features a lightweight, efficient design.

\textbf{LWRES \cite{intro-meth2}.} LWRES is a lightweight residual network model. The lightweight residual network with dilated kernels can significantly improve interference recognition performance while only slightly increasing computational complexity.

\textbf{STDN \cite{baseline-dscnn}.} STDN introduces an optimization method for the soft shrinkage threshold ratio coefficient, enabling adaptive evaluation of the noise level. When combined with the deep convolutional neural network, it can significantly enhance interference recognition performance.

We further analyze the effects of geometric transformations (only-GT) and modal augmentation (only-MA) within the composite data augmentation framework on interference recognition under limited, imbalanced datasets.

\begin{table}[tb]
\vspace{-0.5cm} 
\centering
\caption{Performance of wireless interference recognition by different methods under different data augmentation strategies}
\label{tab:result}
\resizebox{\linewidth}{!}{%
\begin{tabular}{c ccc ccc ccc ccc} 
\toprule
\multirow{2}{*}{\textbf{Method}} 
& \multicolumn{3}{c}{\textbf{Limited}} 
& \multicolumn{3}{c}{\textbf{only-GT}} 
& \multicolumn{3}{c}{\textbf{only-MA}} 
& \multicolumn{3}{c}{\textbf{Composite}} \\
\cmidrule(lr){2-4} \cmidrule(lr){5-7} \cmidrule(lr){8-10} \cmidrule(lr){11-13}
& \textbf{OA}$\uparrow$ & \textbf{Recall}$\uparrow$ & \textbf{KA}$\uparrow$ 
& \textbf{OA}$\uparrow$ & \textbf{Recall}$\uparrow$ & \textbf{KA}$\uparrow$ 
& \textbf{OA}$\uparrow$ & \textbf{Recall}$\uparrow$ & \textbf{KA}$\uparrow$ 
& \textbf{OA}$\uparrow$ & \textbf{Recall}$\uparrow$ & \textbf{KA}$\uparrow$ \\
\midrule
SENET & 0.7663 & 0.7887 & 0.7386 & 0.8543 & 0.8725 & 0.8357 & 0.7548 & 0.7754 & 0.7231 & 0.9125 & 0.9359 & 0.9049 \\
LWRES & 0.7715 & 0.8021 & 0.7851 & 0.8743 & 0.9162 & 0.8594 & 0.7830 & 0.8121 & 0.7725 & 0.9352 & 0.9481 & 0.9246 \\
STDN  & 0.8235 & 0.8230 & 0.8021 & 0.9249 & 0.9212 & 0.9153 & 0.8527 & 0.8447 & 0.8474 & 0.9514 & 0.9479 & 0.9394 \\
\textbf{DiffWIR (Ours)} & \textbf{0.8247} & \textbf{0.8831} & \textbf{0.8020} & \textbf{0.9290} & \textbf{0.9527} & \textbf{0.9196} & \textbf{0.8472} & \textbf{0.8490} & \textbf{0.8306} & \textbf{0.9572} & \textbf{0.9613} & \textbf{0.9462} \\
\bottomrule

\end{tabular}%
}
\vspace{-0.5cm} 
\end{table}

\begin{figure}[tb]
    \centering
    \includegraphics[width=0.6\linewidth]{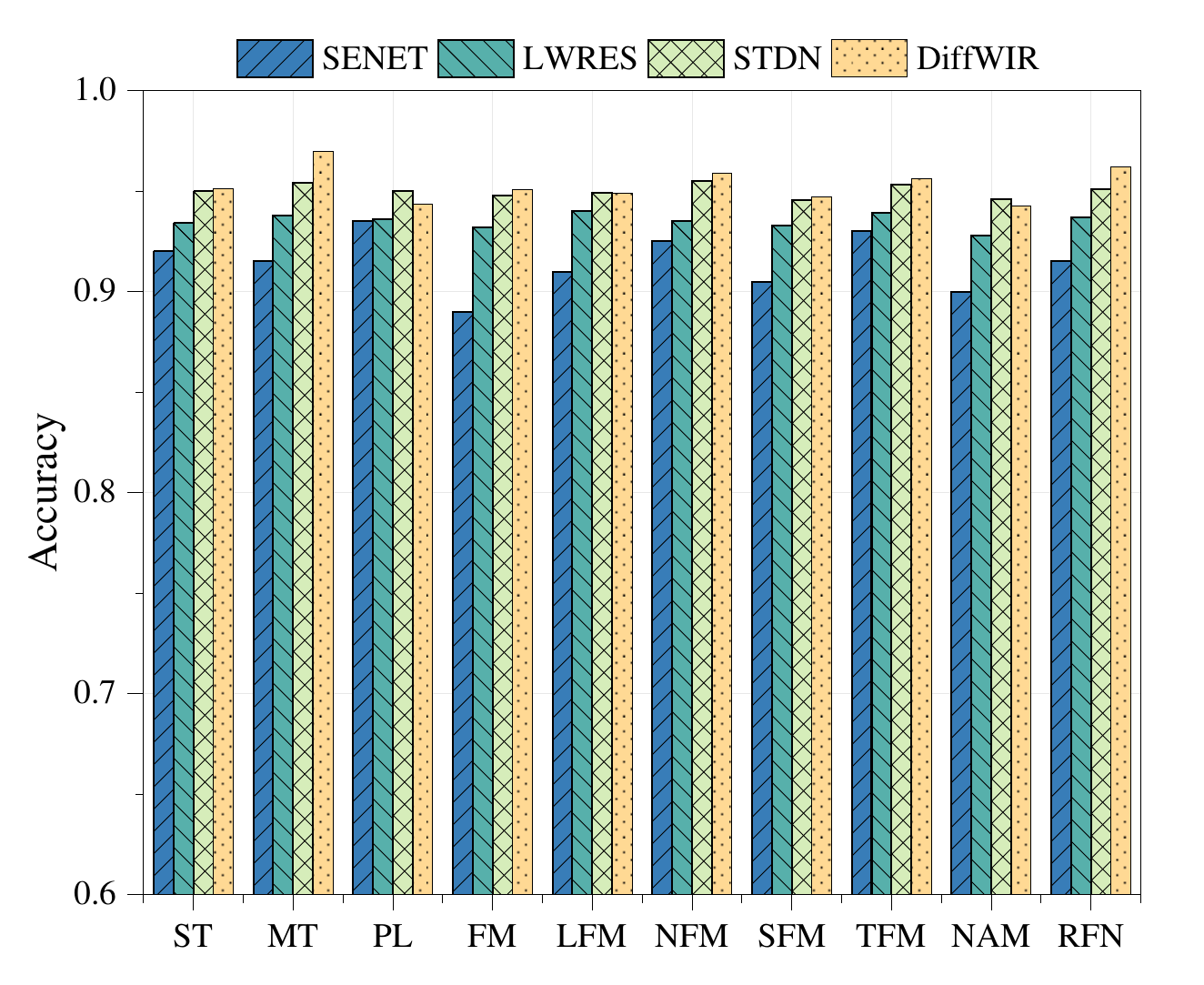}
    \vspace{-2mm}
    \caption{The recognition accuracy of various methods under the composite data augmentation strategy for each interference type is analyzed. Our proposed DiffWIR demonstrates superior performance compared to other methods.}
     \vspace{-2mm}
    \label{fig:final}
\end{figure}

\subsection{Results and Discussion}
As shown in Table~\ref{tab:result}, the proposed DiffWIR model demonstrates superior performance in interference recognition on the real-world dataset, outperforming all compared algorithms. First, the effectiveness of the proposed composite data augmentation framework is verified through an overall comparison. Under the condition of limited and imbalanced datasets, all methods exhibit poor classification performance. Even under these conditions, the proposed method achieves a maximum performance advantage of 5.84\% over the baseline method. This advantage is reflected not only in higher classification accuracy but also in the balanced recognition of different types of interference. Expanding the dataset through geometric transformations can make the training set sufficient and balanced. Under these conditions, all methods demonstrate improved performance, with gains of up to 10.43\%. Interference recognition performance can also be improved through modal augmentation. However, under limited, imbalanced data, this improvement is limited and may even lead to suboptimal results. This is because, as mentioned above, although modal augmentation can improve the representation capability for different types of interference, achieving significant performance improvements is difficult when the overall decision boundary is ambiguous. Therefore, dataset expansion serves as the foundation of data augmentation, while enriching feature representations is a more fine-grained optimization. The proposed composite data augmentation framework integrates the advantages of these two methods. Through the composite augmentation strategy, all methods achieve performance improvements under the original dataset conditions, with gains of up to 15\%. Compared to single geometric transformation data augmentation, the enhanced feature representation provided by modal augmentation improves interference recognition performance, yielding gains of 2\%-7\%.

\begin{figure*}[tb]
    \centering
    \subfloat[Limited]{%
        \includegraphics[width=0.38\textwidth]{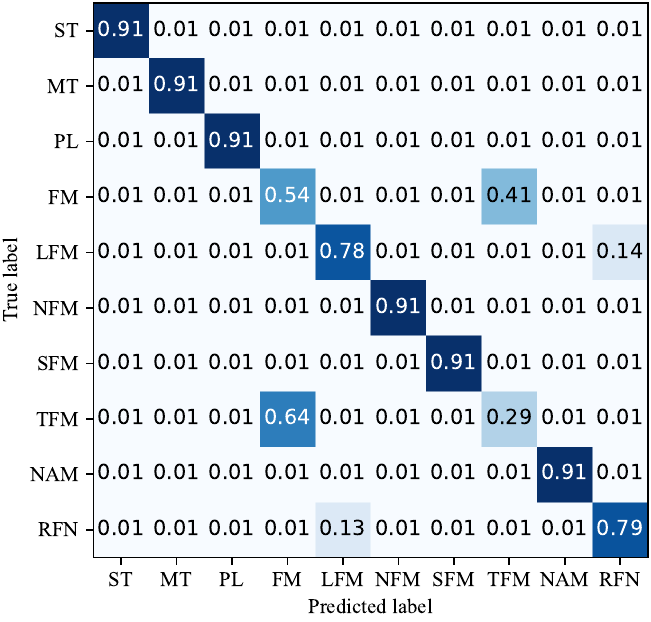}%
    }%
    \hfill 
    \subfloat[only-GT]{%
        \includegraphics[width=0.38\textwidth]{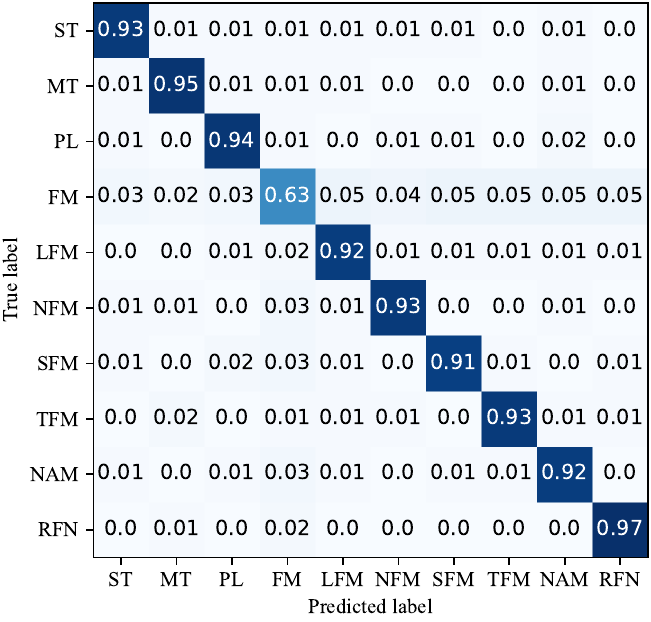}%
    }
    \\ 
    \vspace{-0.3cm}
    \subfloat[only-MA]{%
        \includegraphics[width=0.38\textwidth]{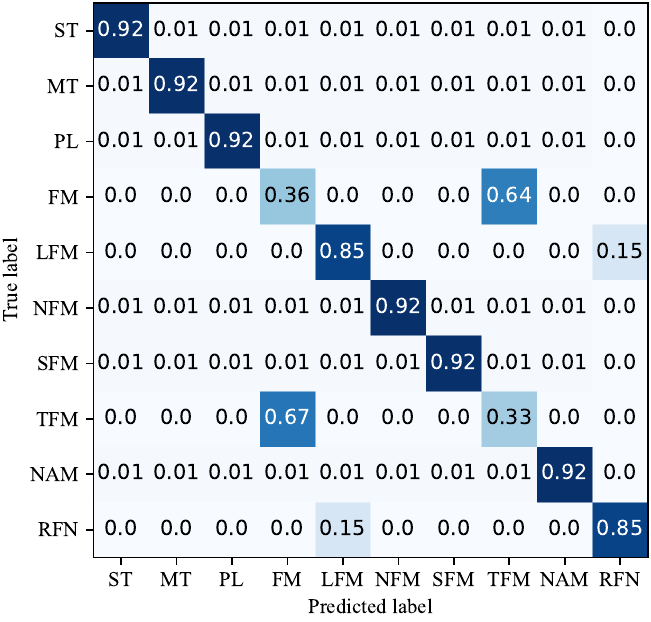}%
    }%
    \hfill
    \subfloat[Composite]{%
        \includegraphics[width=0.38\textwidth]{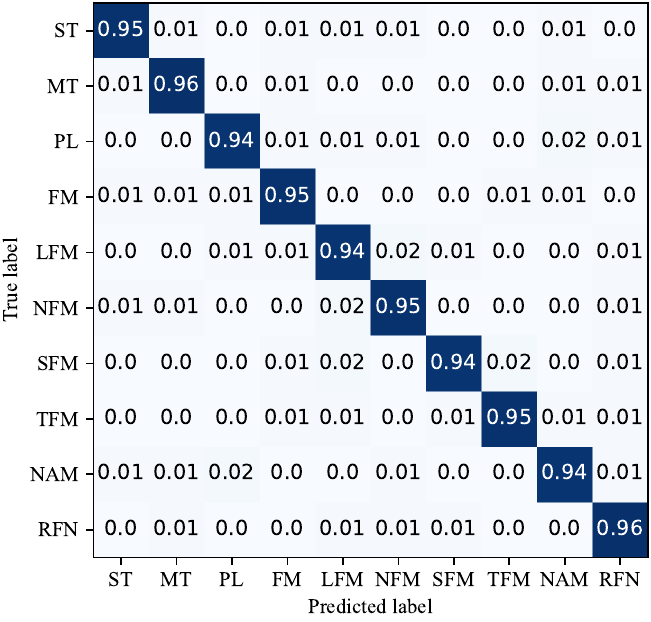}%
    }
    \caption{Confusion matrix describing DiffWIR interference recognition capability under limited data set and various data augmentation methods.}
    \label{fig:cm}
    \vspace{-3mm}
\end{figure*}

Using the composite data augmentation strategy, we analyze the recognition performance of various methods for different interference types. As shown in Fig.~\ref{fig:final}, DiffWIR demonstrates robust overall performance and achieves strong results across most interference types. SENET and LWRES exhibit relatively weak recognition performance because they fail to account for environmental factors. STDN considers noise, thereby improving recognition accuracy. For instance, it achieves the highest accuracy in recognizing the PL and NAM interference types. However, a lack of in-depth consideration of environmental factors prevents it from achieving optimal recognition performance. Finally, to conduct a more in-depth analysis of classification performance, we plot the confusion matrices for DiffWIR on challenging datasets and with various data augmentation strategies, as shown in Fig.~\ref{fig:cm}. Under the conditions of limited and imbalanced datasets, overall interference recognition performance remains weak. Insufficient data prevents the model from obtaining sufficient information for effective distinction, leading to misclassification when distinguishing between LFM and RFN, two types of interference with high discriminative potential. A greater degree of confusion occurs between categories that are highly similar, such as TFM and FM. When comparing the two data augmentation techniques—geometric transformation and modal augmentation—modal augmentation fails to effectively alleviate this confusion. Geometric transformation gradually refines the classification boundary toward the optimal solution by expanding the dataset. The proposed data augmentation framework builds on this by extracting more fine-grained information through multi-modality, thereby achieving more effective and balanced interference recognition.

\section{Conclusion}
This paper explores the potential of generative AI in addressing the challenge of wireless interference recognition. To this end, a diffusion model, DiffWIR, is proposed for the recognition task under limited data conditions. The model adaptively expands the dataset and enriches data distribution information under limited data conditions. It further models the distribution of various interferences in the electromagnetic space to accurately identify interference types. Using the wireless communication dataset collected from a real-world environment, DiffWIR outperforms the baseline model.

\section*{Acknowledgments}
This research was supported in part by the National Natural Science Foundation of China under Grant U24A20247 and Grant 62471277, the Young Elite Scientists Sponsorship Program by CAST under Grant ZB2025-293, the Fundamental Research Funds for the Central Universities under Grant 531119200354, and the Future Industry Technology Innovation Special Project of the Key Science and Technology Support Program of Hebei Province under Grant 2050201.

\bibliographystyle{IEEEtran} 
\bibliography{Reference}
%
%
%
%

\end{document}